\documentclass[9pt,twocolumn,twoside]{pnas-new-alt}
\templatetype{pnasresearcharticle} 

\usepackage[utf8]{inputenc}
\usepackage{graphicx}
\usepackage{amssymb,amsmath}
\usepackage{url}
\usepackage{xcolor}
\usepackage{appendix}
\usepackage{cleveref}
\usepackage{xfrac}
\usepackage{soul}

\title{Data-driven dynamical coarse-graining for condensed matter systems}

\author[a b c e *]{Mauricio J. del Razo}
\author[d c]{Daan Crommelin} 
\author[b]{Peter G. Bolhuis}

\affil[a]{Freie Universität Berlin, Department of Mathematics and Computer Science, Berlin, Germany}
\affil[b]{Van ’t Hoff Institute for Molecular Sciences 1090GD Amsterdam, The Netherlands}
\affil[c]{Korteweg-de Vries Institute for Mathematics 1090GD Amsterdam, The Netherlands}
\affil[d]{Centrum Wiskunde \& Informatica, 1098 XG Amsterdam, The Netherlands}
\affil[e]{Dutch Institute for Emergent Phenomena, 1090GL Amsterdam, The Netherlands}
\affil[*]{E-mail: m.delrazo@fu-berlin.de}

\begin{abstract}
Simulations of condensed matter systems often focus on the dynamics of a few distinguished components but require integrating the dynamics of the full system. A prime example is a molecular dynamics simulation of a (macro)molecule in solution, where both the molecules(s) and the solvent dynamics needs to be integrated. This renders the simulations computationally costly and often unfeasible for physically or biologically relevant time scales. Standard coarse graining  approaches are capable of reproducing equilibrium distributions and  structural features but do not  properly include the dynamics. In this work, we develop a stochastic data-driven coarse-graining method inspired by the Mori-Zwanzig formalism. This formalism shows that macroscopic systems with a large number of degrees of freedom can in principle be well described by a small number of relevant variables plus additional noise and memory terms. Our coarse-graining method consists of numerical integrators for the distinguished components of the system, where the noise and interaction terms with other system components are substituted by a random variable sampled from a data-driven model. Applying our methodology on three different systems -- a distinguished particle under a harmonic potential and under a bistable potential; and a dimer with two metastable configurations --  we show that the resulting coarse-grained models are not only capable of reproducing the correct equilibrium distributions but also the dynamic behavior due to temporal correlations and memory effects. Our coarse-graining method requires data from full-scale simulations to be parametrized, and can in principle be extended to different types of models beyond Langevin dynamics.
\end{abstract}

\newcommand\smallX{
	\mathchoice
	{{\scriptstyle\mathcal{X}}}
	{{\scriptstyle\mathcal{X}}}
	{{\scriptscriptstyle\mathcal{X}}}
	{\scalebox{.7}{$\scriptscriptstyle\mathcal{X}$}}
}

\newcommand\smallU{
	\mathchoice
	{{\scriptstyle\mathcal{U}}}
	{{\scriptstyle\mathcal{U}}}
	{{\scriptscriptstyle\mathcal{U}}}
	{\scalebox{.7}{$\scriptscriptstyle\mathcal{U}$}}
}

\graphicspath{ {figs/} }

\let\newmaketitle\maketitle
\let\maketitle\relax
\makeatother

\begin{document}

\maketitle
\ifthenelse{\boolean{shortarticle}}{\ifthenelse{\boolean{singlecolumn}}{\abscontentformatted}{\twocolumn[\newmaketitle \abscontent] 
}}{}

Computer simulations have been extremely powerful in the study of  (soft) condensed matter systems.  Application of molecular dynamics, Langevin or Monte Carlo enables the modeling of phase transitions, material properties, conformational changes in biomolecules, and many other applications \cite{frenkel2001understanding,leimkuhler2016molecular,Karplus2002,Dror2012,Glotzer2002,Dren2009,zhou2022molecular,Praprotnik2008}. In addition to thermodynamical properties, dynamical simulations also give access to time-correlation-dependent properties such as diffusion, viscosity, mean first passage times for reactive events, relaxation and even aging processes. 
While accurate, but still approximate, (classical) atomistic force fields are available for many type of molecules, they 
become computationally very costly for large systems.
One important reason is that the timescale separation between the fundamental timestep in the integration of the equation of motion and the timescale needed to observe the phenomenon of interest can be many orders of magnitude. This renders the computational effort to reach the physically or biologically relevant timescales needed for e.g. nucleation or protein-ligand (un)binding prohibitively large. 

Precisely for that reason a plethora of coarse-grained methodologies have been developed \cite{espanol2004statistical,Praprotnik2008,saunders2013coarse,Marrink2013,Tozzini2005}, e.g. methods have been developed to coarse-grain entire protein domains and subdomains into single effective coarse-grained particles \cite{bereau2009generic,dama2013theory,jin2022bottom,noid2008multiscale,noid2008multiscaleII}. Recently, deep learning approaches have been used to learn coarse-grained free energy functions using force-matching schemes to train the neural networks \cite{chen2021machine,durumeric2023machine,husic2020coarse,wang2019machine}. In many of these works, coarse-grained force fields have been developed, in which fast but unimportant degrees of freedom have been integrated out, thus reducing the number of degrees of freedom in the model to the most relevant. 

Similar to the original atomistic force field, coarse-grained force field models can be used in molecular dynamics setup, usually Langevin Dynamics, but are much more efficient due to the reduced number of degrees of freedom and the allowed larger time step. In the resulting dynamics, the effect of the integrated-out degrees of freedom is taken care of by a heat bath. 
However, a major drawback is that one no longer resolves the dynamics on short timescales, which are often still needed to reproduce the correct behaviour on longer timescales of (time-correlated) observables.

To recover this time-correlated behaviour, inertial Langevin dynamics or overdamped (Brownian) dynamics are usually performed with an {\em effective} friction or diffusion constant, which also governs the stochastic noise represented by the stochastic Wiener process. However, the choice of the diffusion or friction constant is rather ad hoc, might be position dependent, or even incorporate hydrodynamic interactions \cite{delong2015brownian,ermak1978brownian,sadeghi2020large}. 

From a first principles point of view, it is desirable that a coarse-grained force field can reproduce the correct dynamics. 
Formally, the well-known Mori-Zwanzig formalism achieves precisely that,  
by constructing a generalized Langevin equation in which all the dynamics of the degrees of interest are condensed into three components: the projected dynamics, the so-called memory term, and a noise term   \cite{hijon2010mori,lei2016data,zwanzig1973nonlinear,zwanzig2001nonequilibrium}. In traditional approaches, the dynamics are projected into slow variables. However, it is theoretically possible to perform exact dynamical coarse-graining even in systems without time scale separation by using a wise choice of collective variables, which are not necessarily slow \cite{lu2014exact}.

In essence, the Mori-Zwanzig formalism shows that macroscopic systems with a large number of degrees of freedom can be well described by a small number of relevant variables, with their dynamics given in terms of the three components mentioned above.
However, the actual derivation and practical implementation of reduced (coarse-grained) models following this formalism are often very difficult due to the challenge of accurately computing the memory kernel \cite{grogan2020data,lei2016data,li2015incorporation,luchi2022coarse}. To make this more tractable, approximations such as the short memory approximation are frequently used. The assumptions underlying such approximations can be reasonable for certain systems, but less so for others, especially if there is no clear separation of timescales between the resolved and unresolved degrees of freedom  \cite{ma2016derivation}. In another related work \cite{she2023data}, instead of trying to reproduce the memory kernel, a set of non-Markovian features and the extended dynamic equation are learned by matching the evolution of the correlation functions.

Here, we take a different, data-driven, route. Instead of aiming to derive the various terms in the generalized Langevin equation (GLE), assisted by invoking certain approximations or by using data to fit e.g. an assumed functional form of the memory kernel, or by matching correlation functions, we use a database to sample a noise process that combines the memory term and the noise term of the GLE. In particular, we explore the application of the data-driven heat bath approach developed in \cite{verheul2016data}. Building on earlier work on data-driven stochastic modeling of unresolved processes \cite{crommelin2008subgrid}, this general data-driven methodology was developed and subsequently applied to multiscale simulations of ocean flow in \cite{verheul2017covariate}. 
This is an example of the ongoing development in which data are used to inform or augment physics-based models.
Other examples are the use of QM data to inform molecular dynamics based on electronic structure and the deployment of data-driven closures in computational fluid dynamics \cite{beck2019deep,maulik2019subgrid}. 

In this work, we develop and apply this data-driven approach for molecular condensed matter systems, in particular of distinguishable solute particles dissolved in a bath of fluid particles. This situation applies to (a) complex solute molecule(s) suspended in a solvent, e.g. proteins in water, or polymer solutions.  
Our coarse-graining method integrates out the fluid particles and replaces their influence with a stochastic data-driven model while retaining all time-correlations correctly. This results in a set of numerical integrators for the distinguished components of the system, where the noise and interaction terms are substituted by a random variable sampled from a data-driven model. 
We demonstrate the validity of our methodology on three different systems, all with a three dimensional solvent: 
As a first test case, we consider a single  distinguished particle immersed in a (solvent) bath of fluid particles, with the distinguished particle experiencing an (external) harmonic potential, 
The second system allows the same particle to hop between two states in an external bistable potential, so that long-time correlations become important. The third test case is a constrained dimer with two metastable configurations.

We find that in most cases the coarse-grained model best reproduces the time (auto) correlations and mean first passage times of the fully resolved model when the data-driven model for the auxiliary variable (i.e., the variable that represents the influence of the solvent particles) is made to be dependent on the current velocity of the distinguished particle, as well as on the auxiliary variable at both the current and the previous time step.  However, in the last example, we show that more sophisticated systems might require a more complex set of variables.

The resulting coarse-grained models not only have all the advantages of coarse-grained force fields, i.e. they reproduce the correct equilibrium distributions with a significant speed up w.r.t the fully resolved model, but also correctly predict the dynamic behavior due to memory effects.
However, they still require data from full-scale simulations to construct the data-driven model. 

Our application of the data-driven coarse-grained dynamics methodology enables the prediction of kinetic observables in a bottom-up manner and can in principle be extended to different types of models beyond Langevin dynamics. E.g., it could be included in reaction-diffusion dynamics \cite{hoffmann2019readdy}, their generalized dynamics \cite{del2022probabilistic,del2023chemical} and their multiscale schemes \cite{del2018grand,del2021multiscale,dibak2018msm,kostre2021coupling}; as well as in biomembrane simulations \cite{cooke2005solvent,sadeghi2018particle,sadeghi2020large,wang2010systematically,wang2010systematic}. Other potential applications include computational fluid dynamics and climate modeling \cite{berner2017stochastic,verheul2017covariate,crommelin2008subgrid,beck2019deep,maulik2019subgrid}.
Future work should include testing the method on more complex systems such as proteins, colloids and surfactant solutions. The sampling of the auxiliary variables in high-dimensional cases such as these ones could in principle be handled by implementing a deep learning (neural network) pipeline.

\section{Data-driven dynamical coarse-graining}
\label{sec:ddcg}

\subsection{Separating the distinguished from solvent components}
\label{sec:langevindynamics}
To introduce our data-driven dynamical coarse-graining approach, first consider $L+N$ interacting particles and assume the dynamics of each particle $i$ follows a Langevin equation 
\begin{align}
    \dot{\smallX_i} &= \smallU_i , \notag \\
    m_i\dot{\smallU_i} &=  - \Gamma \smallU_i -\nabla_{i} V(\smallX) + \xi_i(t),
    \label{eq:langevinAll}
\end{align}
where $\smallX_i,\smallU_i$ corresponds to the position and velocity of the $i$th particle, $V(\smallX)$ is the interaction potential, $\Gamma$ the friction coefficient, and $m_i$ the particle's mass. Note that the friction coefficient can be in general a tensor, e.g. for anisotropic particles, but here we take a scalar for simplicity. The noise term, represented by the Wiener process, satisfies the fluctuation-dissipation relation  $\langle \xi_i(t)\xi_j(t') \rangle = 2\Gamma k_BT\delta_{ij}\delta(t-t')$.

We further assume that from these $L+N$ particles, the first $L$ correspond to distinguished particles, while the remaining ones ($N$) correspond to solvent particles. We are mainly interested in the dynamics of distinguished particles. To distinguish them from solvent particles, we will denote their positions and velocities as $x_j,v_j$ with $j=1,\dots ,L$. Similarly, for the bath particles, we will use $q_k,u_k$ with $k=1,\dots N$. Thus the positions and velocities of all the particles in the system are represented by $\smallX=(x,q)=(x_1,\dots, x_L,q_1,\dots,q_N)$ and $\smallU=(v,u)=(v_1,\dots,v_L,u_1,\dots,u_N)$, respectively. To simplify notation, we will further assume all distinguished particles have the same mass $M$ and all the solvent particles have the same mass $m$.  

It is also convenient to split the (pair) potential function $V$ into two parts. The first part, $U$, includes interactions among pairs of distinguished particles, or between an external potential and distinguished particles. The second one, $U_s$, gathers all the pair interactions involving the solvent, thus
\begin{align}
    V(x,q) = U(x) + U_s(x,q).
    \label{eq:potsplit}
\end{align}

We assume that the distinguished particles move on longer time scales than the bath particles. Usually this is the case if the distinguished particles are larger objects or macromolecules (e.g. proteins), and the bath particles are small e.g. solvent molecules like water. 
Inspired by the Mori-Zwanzig formalism \cite{zwanzig2001nonequilibrium} and based on \cref{eq:langevinAll}, we can write the Langevin equation for the $L$ distinguished particles by condensing all the small scale features arising from the interaction with the solvent and the noise into an auxiliary variable $r$,
\begin{align}
    \dot{x_j} &= v_j, \notag\\
    M \dot{v_j} &=  -\Gamma v_j - \nabla U(x) + r. \label{eq:exactDCG}
\end{align}
These equations only describe the dynamics of the distinguished particles; if one knows $r$, one does not need to integrate the solvent dynamics. The $r$ variable captures part of the projected dynamics, the memory kernel and the noise term. In general, $r$ is dependent on $(x,v)$, the positions and momenta of the distinguished particles. If we knew the exact model for $r$ as suggested by the generalized Langevin equation, this equation would correspond to an exact dynamical coarse-graining of the original dynamics (\cref{eq:langevinAll}) \cite{lei2016data, zwanzig2001nonequilibrium}. However, in most cases, we do not know it explicitly, but we can nonetheless try to approximate it from data. This way we do not need to integrate the dynamics of the solvent particles, greatly reducing the computational effort. In other words, we would like to obtain a reduced model of the form
\begin{align}
    \dot{\tilde{x}}_j &= \tilde{v}_j, \notag \\
    M\dot{\tilde{v}}_j &=  -\Gamma \tilde{v}_j - \nabla U(\tilde{x}) + \tilde{r}, 
    \label{eq:coarsemodel}
\end{align}
where we denote with a tilde the reduced model variables. The coordinates of the distinguished particles act as coarse-grained coordinates of the whole system. The random variable $\tilde{r}$ should then be sampled from a data-driven model, which we need to develop. To tackle this, we will first translate the problem into a discrete-time setting and then we will formulate a data-driven numerical integrator for the reduced model.

\subsection{Langevin integrator}
\label{sec:langevinInts}

\begin{figure}[bt]
	\centering
	\includegraphics[width=0.5\textwidth]{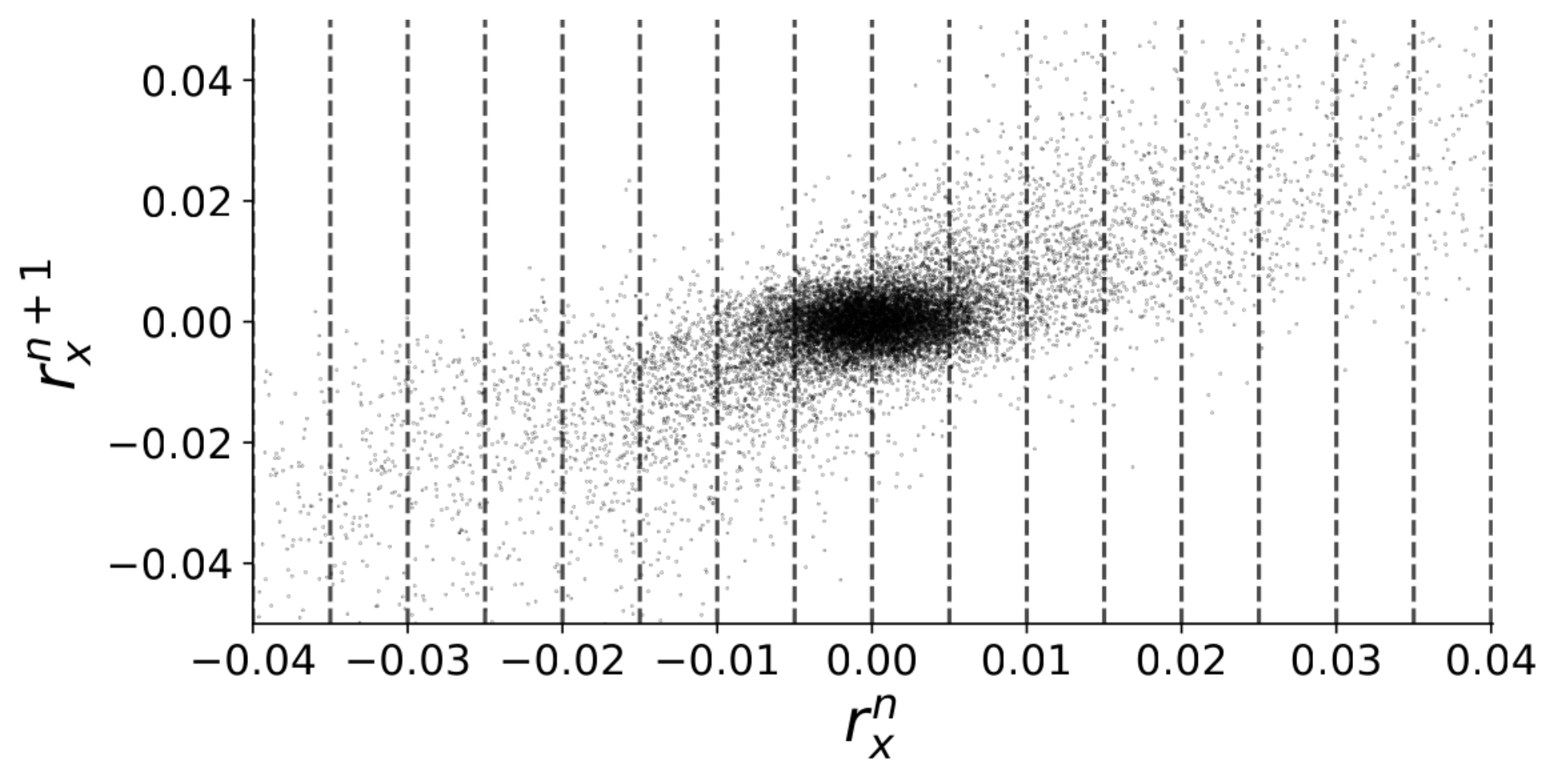}
	\caption{One dimensional binning example fro $r_x^{n+1}|r_x^n$. The cloud of data points from the simulation (\cref{sec:bistablePot}) are projected onto two dimensions: the $x$ coordinate of $r^n$ and the $x$ coordinate of $r^{n+1}$. If $r^{n}$ takes a value inside some bin, then $r^{n+1}$ is sampled uniformly from that bin. The binning implementations in this work are done in between three to nine dimensions, depending on the number and dimension of the conditioning variables.}
	\label{fig:binning}
\end{figure}

Langevin integrators usually rely on splitting methods to improve the accuracy and stability of the integration schemes. These methods are constructed by first decomposing the differential equations into parts that can be solved exactly and then set together a sequence of updates corresponding to an exact solution of each piece in a given time step or fraction of time step.

To obtain and implement a coarse-grained data-driven integrator it is convenient to use a Langevin integrator that can, despite the splitting, integrate a full time-step of the velocity in a single operation. In \cref{app:langevinIntegrators}, we introduce two of the most used Langevin integrators \cite{leimkuhler2016molecular} for an arbitrary Langevin equation \cref{eq:LangevinSplit}: the \textsf{BOAOB} and the \textsf{ABOBA} integrators. The splitting approach in the former does not allow for a full times-step integration of the velocity at once; however, the integration in the latter one does. Thus, in this work, we will base our derivation on the \textsf{ABOBA} scheme, but the derivation is analogous to any other integrator with a splitting that allows for a full-time-step integration of the velocity at once, such as the Stochastic Position Verlet method \cite{leimkuhler2016molecular} among others. It can further be extended to integrators that do not allow for a full-time-step integration of the velocity (e.g. \textsf{BOAOB}); however, this requires sampling the auxiliary variable more than once per time step, resulting in a more cumbersome implementation.

Following \cref{app:langevinIntegrators}, the \textsf{ABOBA} integrator for the Langevin equation from \cref{eq:langevinAll} can be written as follows
\begin{align}
\begin{split}
    \smallX^{n+\sfrac{1}{2}} &= \smallX^n + \smallU^n \frac{dt}{2}, \\
    \smallU^{n+1} &=c_1\smallU^n  -\frac{dt}{2}\mathcal{M}^{-1}\left(1+c_1\right) \nabla U\left(x^{n+\sfrac{1}{2}}\right)  \\ -&\frac{dt}{2} \mathcal{M}^{-1} \left(1+c_1\right) \nabla U_s\left(x^{n+\sfrac{1}{2}},q^{n+\sfrac{1}{2}}\right) + c_2 \mathcal{M}^{-\sfrac{1}{2}} \zeta^n \\
    \smallX^{n+1} &= \smallX^{n+\sfrac{1}{2}} + \smallU^{n+1} \frac{dt}{2},
    \end{split}
    \label{eq:ABOBAalt}
\end{align}
where the velocity integration was condensed into one step,  $c_1 = e^{-\gamma dt}$, $c_2= \sqrt{k_BT (1-c_1^2)}$,  $\mathcal{M}=\text{diag}(M,\dots,M,m,\dots,m)$, and $\gamma=\Gamma \mathcal{M}^{-1}$ with the corresponding potentials. In general, note that as we are assuming $\Gamma$ is scalar and $\mathcal{M}$ a diagonal matrix, $\gamma$ is also a diagonal matrix and the matrix exponentials are easily defined. Alternatively one can write the integration componentwise without using matrices, or simply assume equal masses and friction coefficients, yielding scalar values. The superindices $n$ are the values of the corresponding variable at time $n dt$ with timestep $dt$, and $\zeta^n$ are iid samples from a $3(L+N)$ dimensional standard normal distribution ($\mathcal{N}(0,1)$). To extract the data and to implement the reduced model integrator with this numerical scheme, we split the velocity integration step in Eq.~\ref{eq:ABOBAalt} as follows
\begin{align}
	\begin{split}
    r^{n+1} &= -\frac{dt}{2} \mathcal{M}^{-1} \left(1+c_1\right) \nabla U_s\left(\smallX^{n+\sfrac{1}{2}}\right) + c_2 M^{-\sfrac{1}{2}} \zeta^n, \\
    \smallU^{n+1} &= c_1 \smallU^n -\frac{dt}{2}\mathcal{M}^{-1}\left(1+c_1\right) \nabla U\left(\smallX^{n+\sfrac{1}{2}}\right)  \ + \ r^{n+1},
    \end{split}
\end{align}
where $r^{n+1}$ represents the small-scale features/interactions. We will use this reformulation of the integrator to extract the data for the data-driven integrator of the reduced model.

\begin{figure*}[bt]
	\centering
	\includegraphics[width=1.0\textwidth]{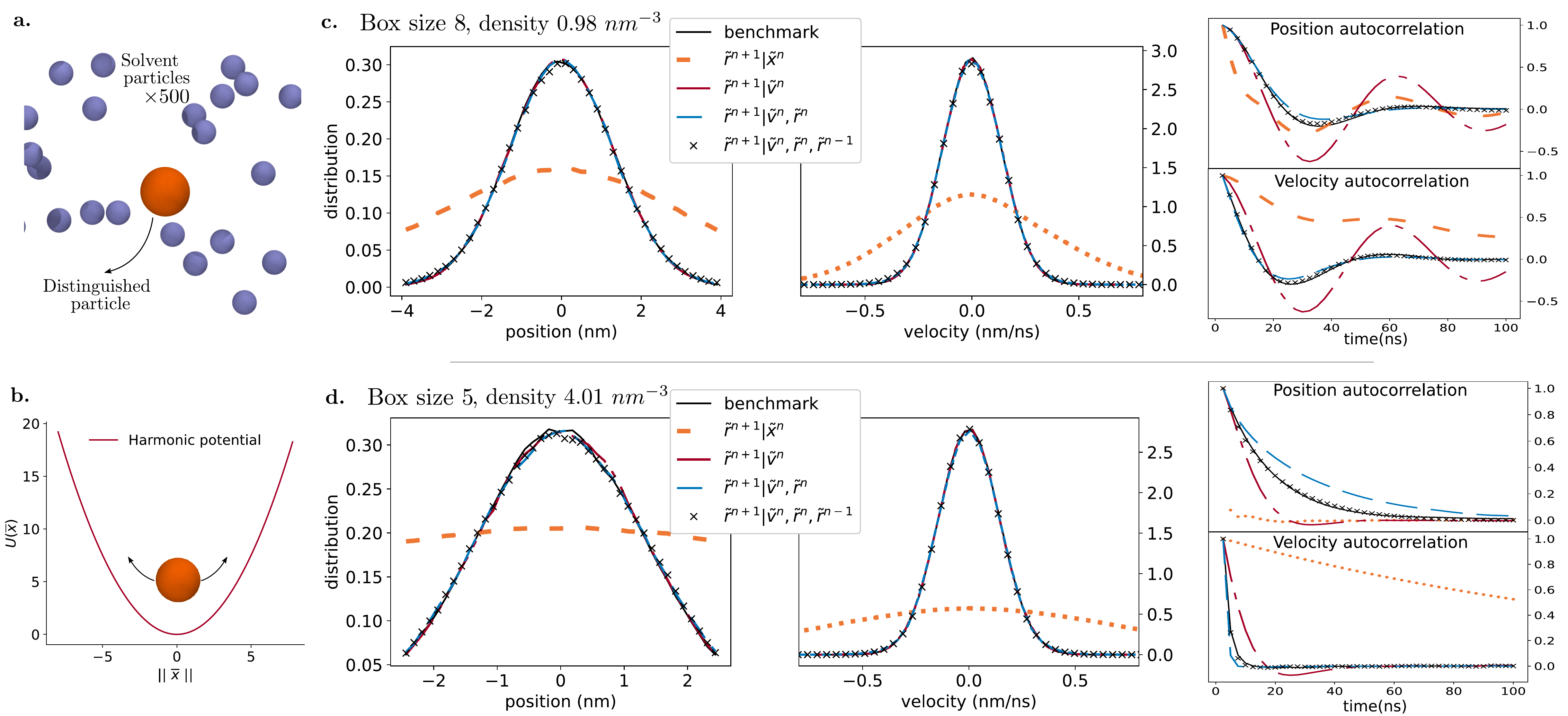} \\
\caption{Harmonic potential example and comparisons of distributions and autocorrelations. a. Illustration of the benchmark simulation. b. Plot of the external harmonic potential (units of $nm$ and $k_B T$) acting on the distinguished particle. c. Comparison of the distributions and autocorrelation functions for the $x$ component of the position and velocity of the distinguished particle between the benchmark simulation and several reduced models with different conditioning. All models use the same parameters and are simulated on a box with an edge length of $8nm$. d. Same comparison but for a box with an edge length of $5nm$.}
	\label{fig:harmonic}
\end{figure*}

\subsection{Data extraction and binning}
\label{sec:databinning}

If we run the Langevin integrator for the full model with the solvent, we can store data points of the form $(t,x,v,r)$, corresponding to time, position, velocity and small-scale features/interactions of the distinguished particle. We do not need to store any of the data from the solvent molecules. For instance at the end of time step $n+1$, we store $(t^{n+1},x^{n+1},v^{n+1},r^{n+1})$. Note that even if we do not store $r^{n+1}$, we could still calculate it from $x^{n+1},v^{n+1}$, although some extra effort is required.

Given these data points for many time steps and many trajectories, we can classify the data under several different assumptions. It is reasonable to expect the random value of $r^{n+1}$ to be dependent on previous values of the position, velocity or even of $r$ itself. This means we would like to sample from its conditional distribution, i.e. from the distribution of $r$ conditioned on some chosen variables $\xi$ \, :
\begin{align}
r^{n+1}| \xi^{n}, \xi^{n-1}, \dots
\label{eq:conddistr}
\end{align}
where $\xi^n$ can take any combination of values of $x^n$, $v^n$ and $r^n$. This yields different possibilities: e.g., we can assume it to be conditioned only on the previous position $r^{n+1}|x^n$, or on one of the previous values of $r$, $r^{n+1}| r^n$ or even two of them $r^{n+1}| r^n, r^{n-1}$. Each of these choices will yield a different data-driven model to sample $r^{n+1}$ at each integration time step (see also \cite{verheul2016data}). In the simplest case, we assume the distribution is only conditioned on the previous position or velocity, $r^{n+1}|x^n$ or $r^{n+1}|v^n$. 

To develop the scheme for the reduced model, at each time step $n$ we need to sample (randomly) from the conditional distribution \ref{eq:conddistr}. We do not know this distribution, however we can approximate sampling from it by resampling from the data (i.e., bootstrapping).  To do so we partition the data into bins corresponding to different values of the conditioning variable $\xi$ (\cref{fig:binning}). In this way, given the conditioning variables, we can determine the corresponding bin and resample uniformly from all the available data in that bin. Of course, binning and resampling is simple in one dimension, but the implementation becomes complex  for higher dimensions, and even unfeasible for very high dimension (due to curse of dimension). Moreover, the higher the dimension, the more likely to encounter empty bins. In such cases, we simply sample from the closest non-empty bin.

\subsection{Data-driven Langevin integrator}
\label{sec:langevinDataInt}
Assuming we have a data-driven model to sample $r^{n+1}$, as the one described in the previous section, we can write the data-driven integrator for the distinguished particles in Eq.~\ref{eq:ABOBAalt}  as follows 
\begin{align}
\begin{split}
\tilde{x}^{n+\sfrac{1}{2}} &= \tilde{x}^n + \tilde{v}^n dt/2, \\
\tilde{v}^{n+1} &= e^{-\gamma dt} \tilde{v}^n -\frac{dt}{2m}\left(1+e^{-\gamma dt}\right) \nabla U\left(\tilde{x}^{n+\sfrac{1}{2}}\right) + \tilde{r}^{n+1}, \\
\tilde{x}^{n+1} &= \tilde{x}^{n+\sfrac{1}{2}} +\tilde{v}^{n+1} dt/2.
\end{split}
\end{align}
Note that as in this case we are assuming equal masses, $\gamma=\Gamma/M$ is simply a scalar. The structure of the integrator follows the same form of the \textsf{ABOBA} scheme, but it samples part of the velocity integration from the chosen data-driven model ($\tilde{r}^{n+1}|\tilde{\xi}^n,\tilde{\xi}^{n-1},\dots $). It can thus integrate the dynamics of the distinguished particle without (explicitly) integrating the bath particles, i.e. we can write everything in terms of only the coarse-grained variables $x$. We refer to these models as reduced models. Note we wrote $\tilde{r}$ instead of $r$ to emphasize that these variables are resampled from the data-driven model and that they belong to the reduced model.

\begin{figure*}[bt]
	\centering
	\includegraphics[width=\textwidth]{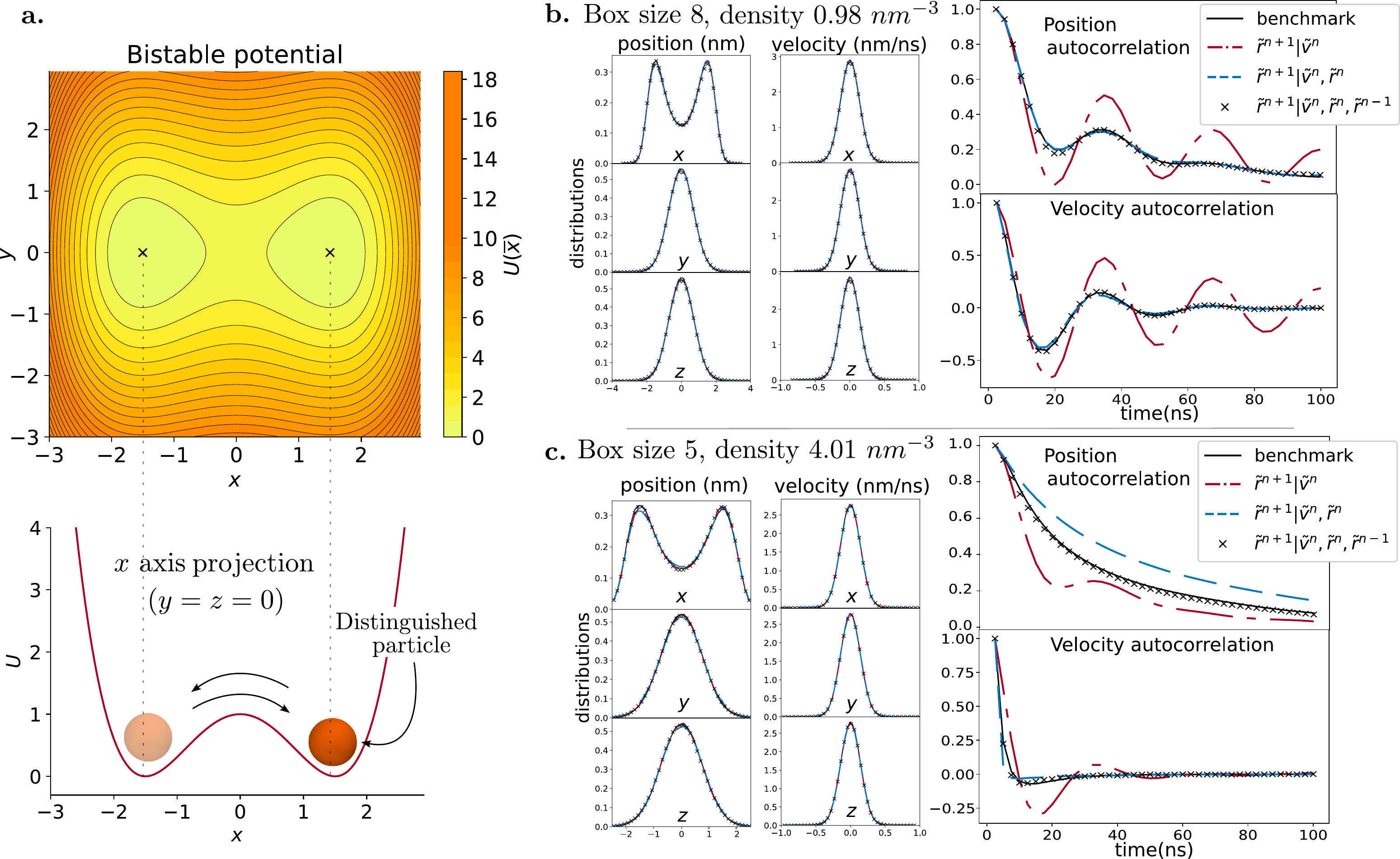}
\caption{Bistable potential example (three dimensional) and comparisons of distributions and autocorrelations. a. Plot of the external bistable potential (units of $k_B T$) acting on the distinguished particle. The potential is axis-symmetric along the $x-axis$.  b. Comparison of the distributions for position and velocity of the distinguished particle between the benchmark simulation and several reduced models with different conditioning, as well as comparisons of the $x$ component of the autocorrelation functions of the position and velocity. All models use the same parameters and are simulated on a box with an edge length of $8nm$. c. Same comparison but for a box with an edge length of $5nm$.}
	\label{fig:bistable}
\end{figure*} 

\section{Reduced models for solvents with WCA interaction}
\label{sec:applicationmodels}

We apply our dynamic coarse-graining method to produce a reduced model of one or two distinguished particles immersed in a solvent in three dimensions. To avoid spurious effects like phase transitions we consider the case in which all pair interactions ---between pairs of solvent particles and between a distinguished particle and a solvent particle--- are mediated by repulsive WCA potentials \cite{andersen1971relationship}. The WCA potential is simply a Lennard-Jones potential truncated and shifted at its minima. This truncation removes the attractive part leaving only a repulsion potential between particles. The WCA potential between the ith and jth particle is given by
\begin{align}
    U_s(\zeta_{ij}) = 
\begin{cases}
      4\epsilon \left[ \left(\frac{\sigma}{\zeta_{ij}}\right)^{12} -
      \left(\frac{\sigma}{\zeta_{ij}}\right)^6 \right] +\epsilon & \text{if } \zeta_{ij}\leq 2^{1/6}\sigma\\
      0 & \text{otherwise}, \\
\end{cases} 
\end{align}
where $\zeta_{ij}$ is the distance between the particles. In addition to the pair interactions, we either incorporate an external potential that only acts on the distinguished particle or a pair potential between distinguished particles. In what follows, we show the results of the reduced model for two external potentials: harmonic and bistable, as well as for a pair potential for two distinguished particles modeling a two-state dimer. For all the simulations below, the parameters of the WCA potential are $\epsilon=1$ and $\sigma = d * 2^{-1/6}$, where $d$ is the cutoff distance of the potential and in this work corresponds to the particles' diameter.

\subsection{Harmonic potential}
\label{sec:harmonicpot}
In this example, we incorporate an external harmonic potential acting only on the distinguished particle in three dimensions, 
\begin{align}
U(x) = \frac{k}{2}  (x\cdot x).
\end{align}
The integration of the full model (including the solvent) is done with the \textsf{ABOBA} scheme from \cref{eq:ABOBAalt}. Following the setup proposed in \cref{sec:langevinInts},  we obtain trajectories of the form $t^n,x^n,v^n,r^n$ corresponding to the time, position, velocity and small scale features of the distinguished particle at time $t_n=n dt$ with timestep $dt=0.05ns$. To generate the data-driven model, we produce 2500 trajectories of the full model, each with a time length of $500 ns$. 

The parameters of the simulation are the following: $N=500$ solvent particles with a diameter of $d=0.5nm$ and a mass of $18g/mol$ (approximately the mass of a water molecule). The friction coefficient is $\Gamma=0.3 (g/mol)/ns$. As we use reduced energy units, we set $k_B T = 1$. The distinguished particle ($L=1$) is assumed to have three times the mass of the solvent particles ($54g/mol$). All the simulations employ periodic boundary conditions and are done on two simulations boxes, one with an edge length of $8nm$ (solvent density $0.98 nm^{-3}$) and another one with $5nm$ (solvent density $4.01 nm^{-3}$). Finally, the harmonic potential uses $k=0.6$.

Using the data produced with the benchmark model, we binned the data following \cref{sec:databinning} using $10$ bins per dimension, and we derived several data-driven models for the solvent interaction by conditioning on different variables. Using each of these data-driven models, we construct reduced models for the distinguished particle following \cref{sec:langevinDataInt}.
For this example, we produce four data-driven models, each of them using different conditioning to sample the small-scale features: $\tilde{r}^{n+1}|\tilde{x}^n$;  $\tilde{r}^{n+1}|\tilde{v}^n$; $\tilde{r}^{n+1}|\tilde{v}^n,\tilde{r}^n$ and $\tilde{r}^{n+1}|\tilde{v}^n,\tilde{r}^n, \tilde{r}^{n-1}$ (\cref{fig:harmonic}). Note that in this example, $\tilde r^n$ is a 3-dimensional vector, and the same holds for $\tilde x^n$ and $\tilde v^n$. Thus, $\tilde{r}^{n+1}|\tilde{x}^n$ denotes the distribution of a 3-dimensional variable conditioned on a 3-dimensional set of conditioning variables, whereas $\tilde{r}^{n+1}|\tilde{v}^n,\tilde{r}^n, \tilde{r}^{n-1}$ has a 9-dimensional set of conditioning variables.

To validate the reduced model, we produce a benchmark simulation of the full model of $100$ trajectories with the same parameters as above, but with a length of $10\mu s$. For comparison, we produce another $100$ trajectories of the same length and with the same parameters using each of the reduced models. \Cref{fig:harmonic} shows the comparison between position and velocities distributions for two different sizes of simulation boxes. It further shows the comparison between autocorrelations functions between the benchmark model and the different reduced models. The plots show a better match for the models with $\tilde{r}^{n+1}|\tilde{v}^n,\tilde{r}^n$ and $\tilde{r}^{n+1}|\tilde{v}^n,\tilde{r}^n, \tilde{r}^{n-1}$. This is expected as the distinguished particle's velocity determines the frequency and  strength of the collisions against solvent particles, and thus it is  essential to account for this  velocity to reproduce the effective forces.
Moreover, the autocorrelations are more accurately reproduced as more history is included in the conditioning variables, modeling better the expected memory effects. This  is particularly visible in the position autocorrelation for the smaller simulation box (higher density), where only the conditioning $\tilde{r}^{n+1}|\tilde{v}^n,\tilde{r}^n, \tilde{r}^{n-1}$ matches the benchmark simulation. This is not surprising since we expect memory effects to be more significant in more dense solvents.

\subsection{Bistable potential}
\label{sec:bistablePot}

\begin{figure}[bt]
	\centering
	\includegraphics[width=0.5\textwidth]{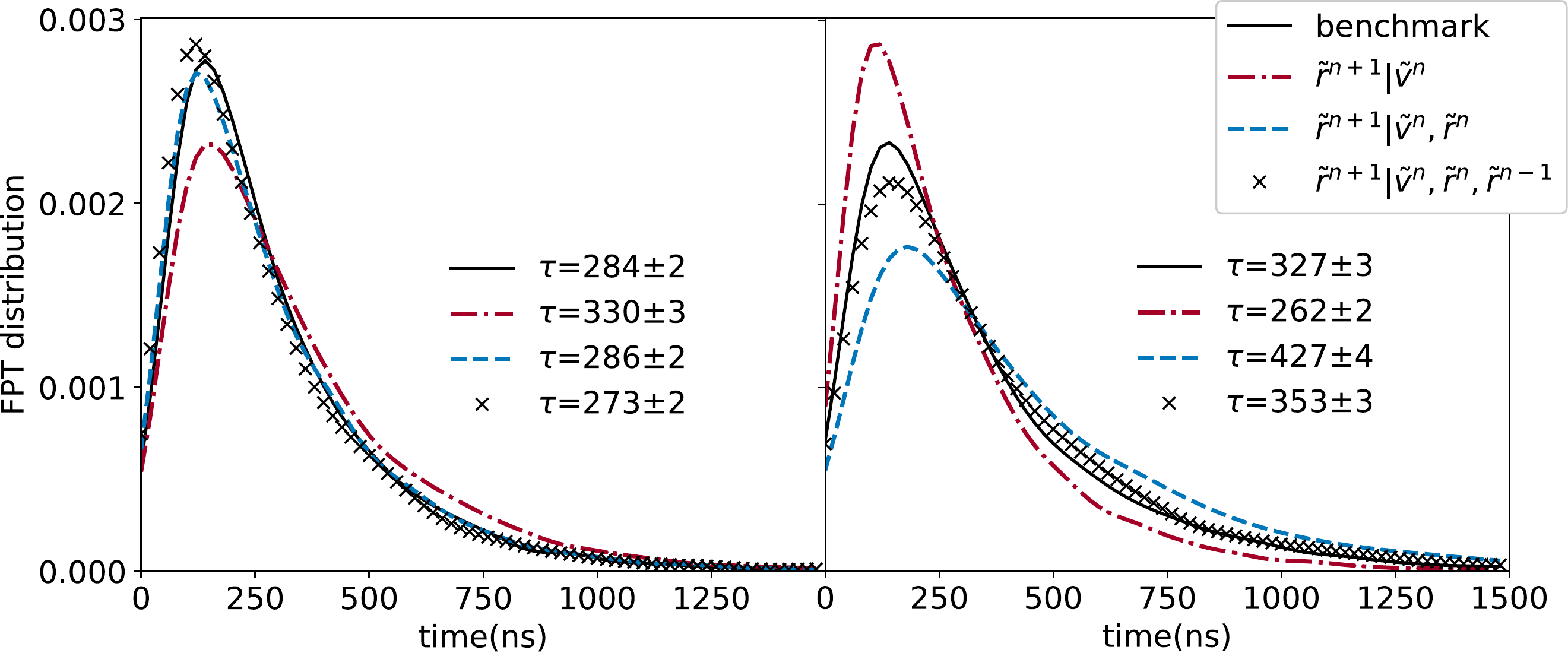} 
\caption{Comparison of the first passage time distributions from one minima to the other for the bistable potential example. Distributions estimated over $10000$ samples for simulations with a box edge length of $8nm$ (left) and $5nm$ (right). Mean first passage times shown in $ns$ calculated over $1000$ bootstrapped samples.}
	\label{fig:bistableFPT}
\end{figure}

Completely analogous to the harmonic potential, we next incorporate  an external bistable potential given by 
\begin{align}
U(\bar{x}) = k\left[ (1-(x/\mu)^2)^2 + y^2 + z^2 \right]
\end{align}
where there is a minima at $(-\mu,0,0)$ and another one at $(\mu,0,0)$. This will provide a more stringent test, as the transition rate will be dependent on the solvent. For the data extraction as well as for the comparison simulations, we use exactly the same parameters as in the previous example, except for the change of the external potential, for which we set $k = 1$, $\mu=1.5$.

For validation, we produce benchmark simulations using the full model, as well as reduced model simulations, both for $100$ trajectories, each with a length of $10\mu s$. In \cref{fig:bistable}, we show comparisons of the position and velocity distributions along cuts through the $x,y$ and $z$ axes, for the reduced models with  $\tilde{r}^{n+1}|\tilde{v}^n$;  $\tilde{r}^{n+1}|\tilde{v}^n,\tilde{r}^n$ and $\tilde{r}^{n+1}|\tilde{v}^n,\tilde{r}^n, \tilde{r}^{n-1}$, respectively. We repeat the simulations for two different simulation boxes with edge lengths $8nm$ and $5nm$ (densities of $0.98$ and $4.01 nm^{-3}$, respectively).

\Cref{fig:bistable} also shows a comparison of the position, velocity and autocorrelation functions for both models, where it is very clear the reduced model with $\tilde{r}^{n+1}|\tilde{v}^n,\tilde{r}^n, \tilde{r}^{n-1}$ is the best to capture the correlations, especially on the more crowded simulations (smaller simulation box). Moreover, for the simulation with an $8nm^3$ box, the autocorrelation function oscillations are due to oscillations of the distinguished particle in the harmonic potential. However, these oscillations are dampened out due to a more dense environment in the simulation with a box of $5nm^3$. 

Finally, \cref{fig:bistableFPT} shows a comparison of the first passage time distributions going from one minima to the other. Note that the mean first passage time increases with density as expected.  Once again the memory effects become relevant in the smaller box, where we need to condition on two past values of the auxiliary variable to accurately reproduce the first passage time distribution.
\begin{figure*}[bt]
	\centering
	\includegraphics[width=0.98\textwidth]{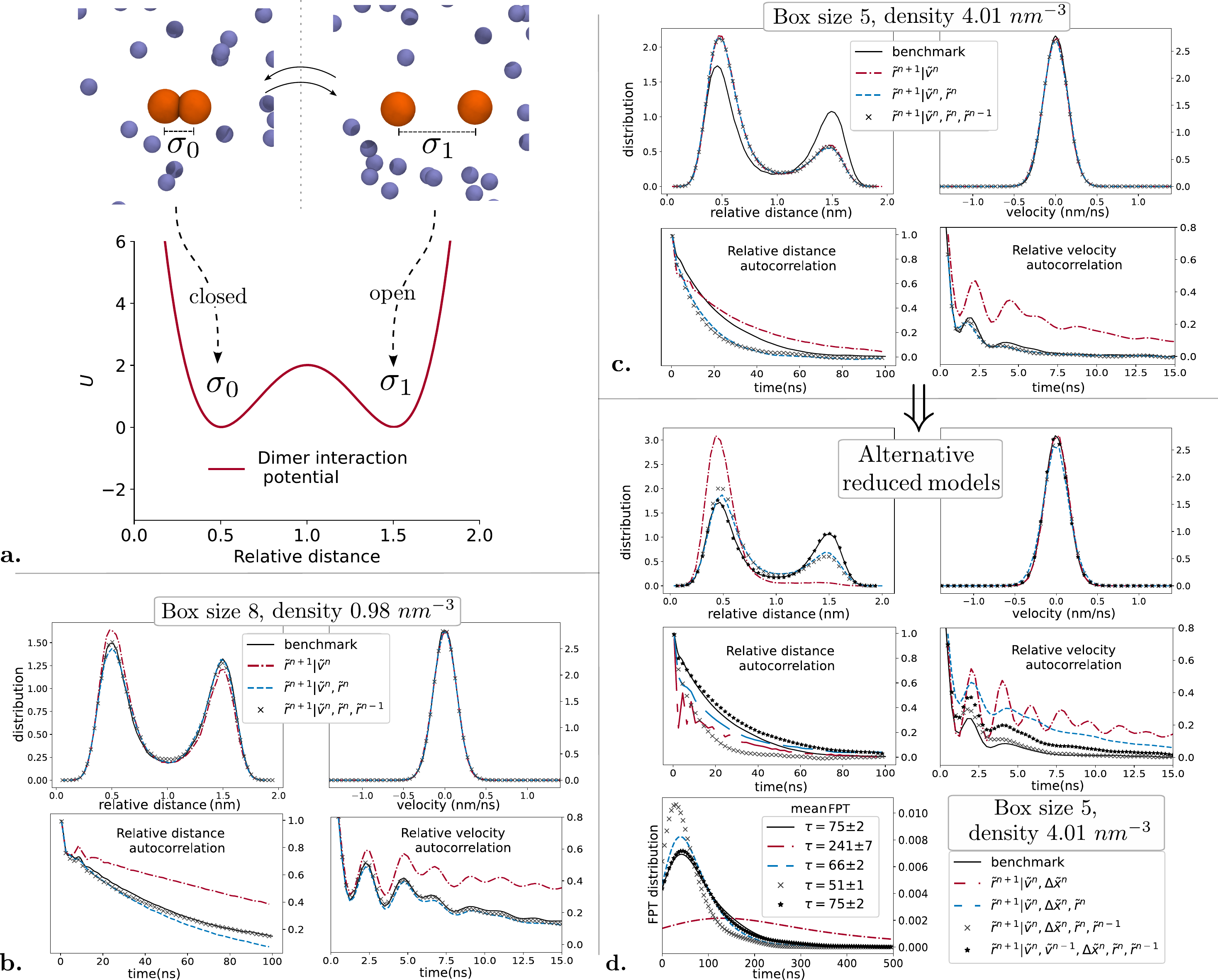}
	\caption{Dimer example (three dimensional solvent) with distributions and autocorrelations comparisons. a. Illustration of the two possible states of the dimer, open or closed, together with the plot of the pair interaction potential (units of $k_B T$). b. Comparison between the benchmark simulation and several reduced models with different conditioning for the following: distribution of the relative distance between dimer particles; distribution of the velocity ($x-$component) of one particle; relative distance autocorrelation; and relative velocity autocorrelation. All models use the same parameters and are simulated on a box with an edge length of $8nm$. c. Same comparison but for a box with an edge length of $5nm$. d. Same comparison as (c) but using alternative variables for the conditioning in the data-driven model. The FPT distributions from the closed state to the open state is also shown, as well as the meanFPTs (ns).}
	\label{fig:pairbistable}
\end{figure*}
\subsection{Two-state dimer}
\label{sec:dimer}

For this example, we consider a dimer of two particles with two metastable configurations determined by the following interaction potential
\begin{align}
U(\Delta x) = 2\left[ 1- \left(\frac{2\Delta x- \sigma_0 - \sigma_1}{\sigma_1 -\sigma_0}\right)^2 \right]^2 
\end{align}
where $\Delta x$ is the relative distance between the particles of the dimer, and we assume $\sigma_0 < \sigma_1$. There is a minima at $\Delta x=\sigma_0$ corresponding to the closed state and another one at $\Delta x=\sigma_1$, corresponding to the open state, both with the same depth (\cref{fig:pairbistable}). Although the simulation is done in three dimensions, the dimer is constrained  to move exclusively along the $x$-axis to enable the efficient testing of different conditioning variables. In this section, we always use $\sigma_0 = 0.5$ and $\sigma_1 = 1.5$. Unless stated otherwise, we once again use the same computational setup and parameters as in the previous section.

\Cref{fig:pairbistable} illustrates the dimer example and compares stationary  distributions and time-autocorrelations between the benchmark model and the reduced models. We compare the distribution of the relative distance between the dimer particles and the distribution of the velocity in $x$-direction of one of the dimer particles, as well as the autocorrelations for the relative distance and $x$-velocity. We do this comparison for two different solvent concentrations (simulation boxes with edge lengths $8nm$ and $5nm$). Because of the constrained motion of the dimer particles, in this example we have dim($\tilde v^n$)=2 (it contains the $x$-velocity component of each dimer particle) and similarly dim($\tilde r^n$)=2.
 
We first note that, by comparing the benchmark distributions for the two simulation boxes in \cref{fig:pairbistable}b and \cref{fig:pairbistable}c, the closed state is more likely to be observed in the smaller box, due to restricted motion in more dense environments. Thus, the dimer state distribution depends not only on the dimer interaction potential but also on the solvent properties and concentration, especially in dense settings. As modeling more dense solvents requires taking into account memory effects, reproducing the distribution precisely will also require modeling the memory accurately in the coarse-grained models.

The comparisons in \cref{fig:pairbistable}b show that the reduced models $\tilde{r}^{n+1}|\tilde{v}^n,\tilde{r}^n$ and $\tilde{r}^{n+1}|\tilde{v}^n,\tilde{r}^n,\tilde{r}^{n-1}$ reproduce the distributions and autocorrelations quite well for the large box (less dense) system. However, as shown in \cref{fig:pairbistable}c, in case of the dense solution the reduced models that use the same conditioning variables give substantial errors in the relative distance distribution and autocorrelation, due to their dependence on the solvent properties. \Cref{fig:pairbistable}d shows that using a broader set of conditioning variables including the relative distance between dimer particles $\Delta \tilde{x}$, we can obtain significantly better results. In particular, the model in \cref{fig:pairbistable}d with $\tilde{r}^{n+1}|\tilde{v}^n,\tilde{v}^{n-1},\Delta\tilde{x}^{n},\tilde{r}^n,\tilde{r}^{n-1}$, reproduces the distributions very accurately as well as the FPT distributions and the mean FPTs. It further yields a better match for the relative distance autocorrelation. However, although still in good agreement, the relative velocity autocorrelation is not as precise as in the example from \cref{fig:pairbistable}b.

This last issue highlights an important aspect of the binning approach: the trade-off between the choice and number of conditioning variables and the dimension of the distribution for the auxiliary variables $r$. While a larger number of conditioning variables likely provides a more complete representation of the relevant dependencies, it also results in fewer data points in the individual bins (assuming the same total amount of data). For instance, in the dimer example, each velocity or $r$ variable used in the conditioning adds two dimensions (one degree of freedom along the $x$ axis per particle), and the relative distance adds one more. Thus, while in the model with $\tilde{r}^{n+1}|\tilde{v}^n,\tilde{r}^n$ the dimension of the set of conditioning variables is four, in the model with $\tilde{r}^{n+1}|\tilde{v}^n,\tilde{v}^{n-1}, \Delta\tilde{x}^{n},\tilde{r}^n,\tilde{r}^{n-1}$ the dimension is nine. Because in the binning approach sparsity grows exponentially with the number of dimensions for fixed amount of data, using more conditioning variables will result in more underpopulated (data-sparse) bins and may lead to inaccuracies in certain regimes, such as we observe in \cref{fig:pairbistable}c and \cref{fig:pairbistable}d. This illustrates a limitation of the binning approach. To better handle high-dimensional settings, it can be replaced with more sophisticated sampling methods, such as the deep-learning method proposed in \cite{crommelin2021resampling}. We leave this for future study. 

Another non-trivial issue arising in more complex systems is: what is the right choice of conditioning variables to construct the reduced model? On one hand, we want that they accurately represent the dependencies of the underlying distribution, but we also want their joint dimension to be as small as possible. We hypothesize that neural network architectures could also be used to find an optimal low-dimensional set of conditioning variables --- perhaps even as linear or non-linear combination of the main variables. In the dimer example, using the relative distance was an obvious choice, but in more complex systems it might not be as straightforward.

\section{Discussion \& Conclusion}

In this study, we developed a coarse-graining method for distinguished particles immersed in a solvent with WCA interactions, inspired by the Mori-Zwanzig formalism and building on earlier work in \cite{verheul2016data}. The numerical integration of the coarse-grained model is based on Langevin numerical integrators, where the interaction and noise terms of the integration step for the distinguished particle are replaced by an auxiliary random variable, avoiding the need to integrate the solvent dynamics. The auxiliary variable is then sampled from a data-driven model, which is constructed based on full-scale simulations; it consists of sampling data from an unknown distribution, which can in principle be conditioned on previous values of the different observables, as well as the auxiliary variable itself. Given an ansatz of possible conditioning variables, we classify and bin the data from the full-scale simulations, so we can approximately sample from the chosen conditional distribution. This results in a data-driven Langevin integrator, which does not require integrating the solvent. As we demonstrate with several numerical examples, it not only accurately reproduces the equilibrium distributions of the positions and velocity but also the dynamics and memory effects, as illustrated in the autocorrelation functions plots. 

Quite remarkably, in the two examples involving transitions between different states of the distinguished particle(s), our method correctly reproduces the long-timescale transition dynamics as characterized by the first passage time (FPT) distributions (cf. figures \cref{fig:bistableFPT} and \cref{fig:pairbistable}d). This is noteworthy as simulations of the Langevin dynamics of only the distinguished particle(s) yield a mean FPT that is off by one order of magnitude.

The main advantage of this approach is the gain in computational efficiency. In all the three examples in this work, the solvent consisted of $N=500$ particles with pair interactions plus the distinguished particle(s). In the reduced models, we do not need to integrate the solvent, and thus the efficiency gain is of the order of somewhere between $N$ and $N^2$, depending on the algorithm employed to calculate the pair interactions. Moreover, once the effects of the solvent are parameterized in the reduced model, we can run the simulations on much larger spatial domains with the same efficiency. This means we can use simulations in smaller domains to construct the reduced model, and then, assuming the same solvent concentration, simulate the systems in larger domains with the reduced model. This is  essential to efficiently implement multi-molecular simulations; albeit 
higher-order correlations between the solvent and  interacting molecules might not be taken into account. This opens the door to new research directions to tackle these issues.

Although our approach is very successful for the simple systems that we investigated, it does have limitations. More complex systems are likely to require a larger set of conditioning variables to capture the dependencies of the distribution of the auxiliary variable $r$, as also discussed in section \ref{sec:dimer}. Thus, parameterizing $r$ in the data-driven model suffers from the curse of dimensionality, because with the binning method used here the data required to successfully parameterize it grows exponentially for each additional conditional variable. One promising approach to sample from a high-dimensional and unknown distribution is to use generative neural network architectures, such as conditional variational autoencoders \cite{doersch2016tutorial} and diffusion models \cite{yang2022diffusion}. Alternatively, neural network resampling \cite{crommelin2021resampling} has been successfully used in the past in similar setups. We leave these endeavors for future work.

An important issue arising in more complex systems is how to choose the conditioning variables. Although physical intuition is indeed helpful, the optimal choice is not trivial. One could simply choose a combination of variables involving positions, velocities and auxiliary variables at the current or previous time step. However, a linear or even a non-linear combination of these variables is perhaps more optimal, such as the relative distance in the dimer case in section \ref{sec:dimer}, or perhaps the angular velocity of a more complex molecule. Similarly, it can be advantageous to use variable values at multiple previous time steps, possibly separated by several time steps. An interesting approach to explore would be to engineer neural network architectures able to find an optimal low-dimensional set of conditioned variables while simultaneously doing the training.

Another point is that in our work we have kept the integration time steps of the full model and the reduced model identical in our work. In standard coarse graining approaches a longer time step can be taken because the effective potentials are usually smoother. Here, this is not the case, as we are still using the bare solute-solute interaction potentials.  Still, as solutes are usually larger and move slower, it is well possible that a longer time-step can be used (as was also done in \cite{crommelin2008subgrid,verheul2016data}, albeit for different systems). This will be investigated in future work.

To conclude, the methodology here presented is not limited to condensed matter systems or to Langevin-type of dynamics. In principle, it can be extended to develop \textit{data-informed physics models} for general complex dynamical systems, where a numerical integrator is employed but not all the degrees of freedom need to be resolved as long as their effect on the variables of interest are represented. Some examples of possible application fields are: climate modeling, social dynamics modeling, biochemical systems modeling, modeling of power, transportation or communication systems as well as general agent-based models. 

\section*{Acknowledgments}
The authors acknowledge support by the Dutch Institute of Emergent Phenomena
(DIEP) cluster at the University of Amsterdam. MJR also acknowledges support from DFG Grant No. RA 3601/1-1.

\bibliographystyle{plain}
\bibliography{references}

\appendix

\noindent 
\section*{APPENDIX}
\section{Langevin Integrators}
\label{app:langevinIntegrators}
The goal of the Langevin integrator is to numerically solve the Langevin equation. For an arbitrary system of $N$ particles in three-dimensional space, with positions $x=(x_1,\dots,x_N)$ and velocities $v=(v_1,\dots,v_N)$ and under a force-field determined by the potential $V(x)$, the Langevin equation in its stochastic differential notation is
\begin{align}
\hspace{-5mm}
\begin{bmatrix}
dx _t\\
\mathcal{M} dv_t
\end{bmatrix} = 
\underbrace{\begin{bmatrix}
v dt\\
0
\end{bmatrix}}_\textsf{A} +
\underbrace{\begin{bmatrix}
0\\
-\nabla V(x) dt 
\end{bmatrix}}_\textsf{B} +
\underbrace{\begin{bmatrix}
0\\
-\Gamma v dt + \sqrt{2k_BT} \ \Gamma^{\sfrac{1}{2}} dW_t,
\end{bmatrix}}_\textsf{O}
\label{eq:LangevinSplit}
\end{align}
where $\mathcal{M}$ is a diagonal matrix with the corresponding masses along the diagonal, $\Gamma$ is the friction coefficient/tensor (inverse mobility) and $W$ is a $3N-$dimensional uncorrelated standard Brownian motion (Wiener process). The splitting into the terms $\textsf{A}$, $\textsf{B}$ and $\textsf{O}$ will be convenient to derive some of the schemes. A detailed account of integrators for the Langevin equation can be found in \cite{leimkuhler2016molecular}. 
It is also convenient to write this equation in terms of the friction coefficient/tensor per unit mass $\gamma = \Gamma \mathcal{M}^{-1}$. In the derivations below, for simplicity, we assume $\gamma$ is a scalar and not a tensor, but the expressions are analogous for the tensor case.

Splitting methods improve the accuracy and stability of the schemes. They require first decomposing the differential equations into parts that can be solved exactly and then set together a sequence of updates corresponding to an exact solution for each piece. \Cref{eq:LangevinSplit} shows a very common splitting, where each of the terms labeled $\textsf{A}$, $\textsf{B}$ and $\textsf{O}$ can be solved exactly when handled independently of each other. Although one would assume the computational cost of these methods to be higher due to the splitting, in practice, most of them only require evaluating the force field once per time step, which in large simulations is by far the most computationally costly part of the integrator.

The implementation details of the methods proposed in this work will depend on the Langevin integrator chosen. Below we show two of the most used integrators and their corresponding advantages and disadvantages in the context of this work. We assume $x^n,v^n$ are the values of $x$ and $v$ at time $n dt$ with timestep $dt$. 

%

\subsection*{\textsf{BAOAB} integrator}
\label{sec:BAOAB}

This scheme is constructed by solving the $\textsf{B}$ and $\textsf{A}$ parts first for half a time step, then the $\textsf{O}$ part for a full-time step, followed by $\textsf{A}$ and $\textsf{B}$ for another half time step. Thus, the implementation of the scheme for one-time step is as follows:
\begin{align}
	\begin{split}
v^{n+\sfrac{1}{2}} &= v^n + \textsf{B}\left(x^n, dt/2\right),   \\
x^{n+\sfrac{1}{2}} &= x^n + \textsf{A}\left(v^{n+\sfrac{1}{2}}, dt/2\right), \\
\hat{v}^{n+\sfrac{1}{2}} &= \textsf{O}\left(v^{n+\sfrac{1}{2}},dt\right), \\
x^{n+1} &= x^{n+\sfrac{1}{2}} + \textsf{A}\left(\hat{v}^{n+\sfrac{1}{2}}, dt/2\right), \\
v^{n+1} &= \hat{v}^{n+\sfrac{1}{2}} + \textsf{B}\left(x^{n+1}, dt/2\right).
\end{split}
\end{align}
with 
\begin{align}
	\begin{split}
    \textsf{B}(x,\tau) &= -\mathcal{M}^{-1}\nabla V(x) \tau, \\
    \textsf{A}(v,\tau) &= v \tau, \\
    \textsf{O}(v,\tau) &= e^{-\gamma \tau} v + \sqrt{k_B T(1-e^{-2\gamma \tau })}\mathcal{M}^{-\sfrac{1}{2}} \zeta^n,
    \end{split}
\end{align}
where $\zeta^n$ is a sample from a $3N$ dimensional standard normal distribution ($\mathcal{N}(0,1)$). This algorithm works well and with much larger time steps than the more simple symplectic Euler method. However, notice the velocity is integrated in steps 1,3 and 5, resulting in a labor-intensive implementation of the reduced methods presented in this work as the auxiliary variable will need to be sampled three times per time step.

\subsection*{\textsf{ABOBA} integrator}
\label{sec:ABOBA}
Alternatively, we can switch the order of the integrators for the different parts obtaining different splitting methods. Here is another popular alternative,
\begin{align}
\begin{split}
x^{n+\sfrac{1}{2}} &= x^n + \textsf{A}\left(v_{i}, dt/2\right), \\
v^{n+\sfrac{1}{2}} &= v^n + \textsf{B}\left(x^{n+\sfrac{1}{2}}, dt/2\right),   \\
\hat{v}^{n+\sfrac{1}{2}} &= \textsf{O}\left(v^{n+\sfrac{1}{2}},dt\right), \\
v^{n+1} &= \hat{v}^{n+\sfrac{1}{2}} + \textsf{B}\left(x^{n+\sfrac{1}{2}}, dt/2\right), \\
x^{n+1} &= x^{n+\sfrac{1}{2}} + \textsf{A}\left(\hat{v}^{n+1}, dt/2\right).
\end{split}
\end{align}
This one is particularly helpful for implementing the reduced models in this work since all the velocity integrations are done successively. We can actually condense all the velocity integration steps into one step, 
\begin{align}
	\begin{split}
    v^{n+1} &= \hat{v}^{n+\sfrac{1}{2}} + B\left(x^{n+\sfrac{1}{2}}, dt/2\right) \\
    &= \textsf{O}\left(v^{n+\sfrac{1}{2}},dt\right) -\frac{dt}{2}\mathcal{M}^{-1} \nabla V\left(x^{n+\sfrac{1}{2}}\right).
    \end{split}
\end{align}
Expanding $\textsf{O}\left(v_{n+\sfrac{1}{2}},dt\right)$, yields
\begin{align}
	\begin{split}
    \textsf{O}\left(v^{n+\sfrac{1}{2}},dt\right) &= \textsf{O}\left(v^n + B\left(x^{n+\sfrac{1}{2}}, dt/2\right),dt\right) \\
    &= e^{-\gamma dt} \left(v^n + B\left(x^{n+\sfrac{1}{2}}, dt/2\right)\right) \\ & \qquad\qquad + \sqrt{k_B T(1-e^{-2\gamma dt})} \mathcal{M}^{-\sfrac{1}{2}}\mathcal{\boldmath{N}}(0,1) \\
    &= e^{-\gamma dt} v^n -\frac{dt}{2} e^{-\gamma dt}\mathcal{M}^{-1} \nabla V\left(x^{n+\sfrac{1}{2}}\right) \\ & \qquad\qquad  + \sqrt{k_B T(1-e^{-2\gamma dt})}\mathcal{M}^{-\sfrac{1}{2}} \mathcal{\boldmath{N}}(0,1).
    \end{split}
\end{align}
Gathering the results
\begin{align}
	\begin{split}
    v^{n+1} &=e^{-\gamma dt} v^n -\frac{dt}{2}\mathcal{M}^{-1}\left(1+e^{-\gamma dt} \right) \nabla V\left(x^{n+\sfrac{1}{2}}\right) \\ & \qquad  + \sqrt{k_B T(1-e^{-2\gamma dt})}\mathcal{M}^{-\sfrac{1}{2}} \mathcal{\boldmath{N}}(0,1).
    \end{split}
\end{align}
Finally, for our implementation, it will be helpful to divide the potential into two terms: one involving the distinguished particles and external potentials $U(x)$ and the other one involving all the interactions with solvent $U_s(x)$ as suggested in \cref{eq:potsplit},
\begin{align}
	\begin{split}
    v^{n+1} &= e^{-\gamma dt } v^n -\frac{dt}{2}\mathcal{M}^{-1}\left(1+e^{-\gamma dt }\right) \nabla U\left(x^{n+\sfrac{1}{2}}\right) \\ &  \qquad -\frac{dt}{2}\mathcal{M}^{-1} \left(1+e^{-\gamma dt}\right) \nabla U_s\left(x^{n+\sfrac{1}{2}}\right) \\ &  \qquad + \sqrt{k_B T(1-e^{-2\gamma dt})}\mathcal{M}^{-\sfrac{1}{2}} \mathcal{\boldmath{N}}(0,1).
    \end{split}
\end{align}

\end{document}